\documentstyle[graphics,epsfig,floats,aps]{revtex}

\begin{document}

\draft
\catcode`\@=11 \catcode`\@=12
\twocolumn[\hsize\textwidth\columnwidth\hsize\csname@twocolumnfalse\endcsname
\title{Electron space charge effect on spin injection
into semiconductors}
\author{Yue Yu$^1$, Jinbin Li$^1$ and  S. T. Chui$^2$}
\address{1. Institute of Theoretical Physics, Chinese Academy of
Sciences, P.O. Box 2735, Beijing 100080, China}
\address{2. Bartol Research Institute,
University of Delaware, Newark, DE 19716}

\date{\today}

\maketitle

\begin{abstract}
We consider spin polarized transport in a
ferromagnet-insulator/semiconductor/insulator-ferromagnet
(F1-I-S-I-F2) junction. We find that the spin current is strongly
dependent on the spin configurations, the doping and space charge
distribution in the semiconductor. When the
ferromagnet-semiconductor interface resistance is comparable to
the semiconductor resistance, the magnetoresistance ratio of this
junction can be greatly enhanced under appropriate doping when the
space charge effect in the nonequilibrium transport processes is
taken into consideration.
\end{abstract}

]

There is much interest recently in incorporating magnetic elements
into semiconductor structures. One of the current focus lies in
injecting spin polarized electrons into non-magnetic
semiconductors\cite{Dat,Mon,Aro,Prin,STMST,mire,hu}. This is
partially motivated by the high magnetoresistance observed in
ferromagnet tunnel junctions\cite{moo}. However, experiments have
so far observed a small magnetoresistance ratio of 1\%
\cite{lee,ham} in F-S-F structures. Schmidt et al \cite{Schm}
emphasized that the large resistance of the nonmagnetic
semiconductor dominates the total resistance and thus lowers the
magnetoresistance ratio. A similar discussion based on a ballistic
picture was given by Grundler \cite{grun} and Hu et al \cite{hu}.
Rashba pointed out that in a ferromagnetic
metal/insulator/semiconductor (FIS) junction the resistance of the
barrier insulator layer can be comparable to that of the
semiconductor, $\Delta R/R$ may be raised by a considerable
percentage \cite{Rash}. Many different geometries of the tunneling
junction were discussed in Ref.\cite{fert}.

Although it is possible that to lift the magnetoresistance to near
10\% with a very high electric field \cite{fla}, there is no a low
field scenario yet. On the other hand, the effect of the Coulomb
interaction between the electrons in the non-magnetic
semiconductor in the nonequilibrium transport process was not
taken into consideration in the above mentioned works. In previous
works on the ferromagnetic metal/insulator/ferromagnetic metal
junction \cite{chui},  one of us found that this interaction
between the electrons in nonequilibrium transport processes played
an important role in the characteristics of the device. Under
steady state nonequilibrium conditions, a magnetization dipole
layer much larger than the charge dipole layer is induced at the
interface. The magnetization dipole layer is zero under
equilibrium conditions. In this paper, we consider the role played
by the Coulomb interaction in an F1-I1-S-I2-F2 junction. To
completely solve the nonequilibrium transport processes with
interactions is highly non-trivial. It requires solving self
consistently Boltzmann type spin transport equation with the
Poisson equation. What we want to touch in this work is using a
simple Hartree approximation to check if the effect comes from the
interactions is significant. We found that the spin polarized
current is strongly affected by the interaction in this
approximation. Because of the large difference in the
spin-dependent current between the parallel and anti-parallel
configurations, which is caused by the space charge effect, the
total resistance becomes dependent on the spin configuration. We
found that magnetoresistance ratio $\Delta R/R$ can be increased
by one order of magnitude higher than the result obtained without
the interactions under appropriate doping of the semiconductor.
Typically, $\Delta R/R\sim 3\%$ for the non-interaction electrons
while $\Delta R/R$ may attain values as high as $\sim 50\%$ due to
the space charge. (See, Figures 2 and 3 below.) We also study
another physical observable, the polarization
$P=\frac{j_+-j_-}{j_++j_-}$ and find that it is strongly dependent
on the space charge distribution. We now describe our result in
detail.

\vspace{2mm}

\noindent{\it General description:} We consider the junction
F1-I1-S-I2-F2. The thickness of the metal (F1, F2), the insulating
barriers (I1, I2) and the semiconductor (S) are denoted by
$L^{L,R}$, $d_{1,2}$ and $x_0$, respectively (See, Figure 1). For
the practical case, $x_0$ is less than the spin diffusion length
in the semiconductor $l_N$. The charge screening lengths in the
metal and the semiconductor are denoted by $\lambda_{L,R}$ and
$\lambda_N$, respectively. In our model, we assume $\lambda_N\ll
l_N$ and $\lambda_{R,L}\ll l_{R,L}$, the spin diffusion lengths in
the metal. Typically, $\lambda\sim 10^{-1}$nm, $l\sim 10^1$nm, and
$l_N> 1\mu$m. $x_0\sim 100$nm to $1\mu$m, depending on the
structure of junctions. The screening length in the semiconductor,
$\lambda_N$, is dependent on the doping of the semiconductor and
can vary in a wide range, say 10nm for the heavy doped
semiconductor and 100nm-1$\mu$m for the lightly doped or undoped
semiconductor.

The problem we would like to solve is defined by the following
four sets of equations.

\noindent (1) The first is the total charge-current conservation,
\begin{eqnarray}
\nabla\cdot {\bf j}=-\frac{\partial \rho}{\partial t},\label{1}
\end{eqnarray}
where $\rho$ is the charge density.

\noindent (2) The second is the diffusion equations of the spin
$s$-dependent current. In order to allow an analytic analysis, we
take a simple Hartree approximation into account to see the space
charge effect. Under such an approximation, the diffusion
equations reads \cite{chui}
\begin{eqnarray}
j_s=\sigma_s(\nabla\mu_s-\nabla W_0+E),\label{2}
\end{eqnarray}
where the magnitude of the electric charge has been set as one;
$E$ is the external electric field; the chemical potential $\mu_s$
is related to the charge density $\rho_s$ by $\nabla\mu_s
=\frac{\nabla\rho_s}{N_s}$ where $N_s$ is the spin-dependent
density of state. $W_0=\int d\vec r' U_{int}(\vec r-\vec r'
)\rho(\vec r') $ is the potential caused by the screened Coulomb
interaction $U_{int}(\vec r)$ and $\nabla W_0$ is the so-called
screening field induced by $U_{int}(\vec r)$. (In the present
device geometry, the three dimensional integration is reduced to a
one-dimensional one.)

\noindent (3) The third is the magnetization relaxation equation
where the magnetization density $M=\rho_\uparrow-\rho_\downarrow$
relaxes with a renormalized spin diffusion length $l$. In the
relaxation time approximation, one has
\begin{eqnarray}
\nabla^2M-M/l^2=0.\label{3}
\end{eqnarray}

\noindent (4) The fourth is the boundary condition at the
interfaces
\begin{eqnarray}
\Delta\tilde\mu_s-\Delta W=r(1-s\gamma)j_s, \label{4}
\end{eqnarray}
where $\tilde\mu_s=\mu_s+Ex$; $Ex$ is the voltage drop on the left
side of the barrier and $\Delta W $ is the electric potential drop
across the barrier, which is assumed much smaller than
$\Delta\mu_s$; $r_s=r(1-s\gamma)$ is the barrier resistance. We
assume that there is no spin relaxation in the insulator, the
spin-dependent currents are continuous across the junctions,
$j^L_s(-d_1/2)=j^N_s(d_1/2)$ and
$j^R_s(x_0+d_2/2)=j^N_s(x_0-d_2/2)$.

In addition, we have the neutrality condition for the total
charges ($Q_{L,N,R}$, e.g., $Q_N=\int^{x_0+d_1/2}_{d_1/2} \rho
dx$) accumulated at the interfaces. By Gauss' law, for the point
$d_1/2<x<x_0+d_1/2$, i.e., inside the semiconductor, the potential
$W_0$ is determined by
\begin{eqnarray}
\nabla W_0(x)= 4\pi Q_L+4\pi\int_{d_1/2}^x\rho dx, \label{5}
\end{eqnarray}
whose constant part of the right hand side gives the constraint on
the charge while the $x$-dependent part gives the function form of
the potential $W_0$. Another constraint is the neutrality of the
system:
\begin{eqnarray}
Q_L+Q_N+Q_R=0. \label{6}
\end{eqnarray}

With those sets of equations (Eqs.(\ref{1})-(\ref{4})) and the two
constraint on the charges (Eqs.(\ref{5}) and (\ref{6})), the
problem can be solved. The formal solutions of the problem are
\begin{eqnarray}
\rho^{L}(x)&=&\frac{\lambda_L}{l_L}\rho^L_{10}e^{(x+d_1/2)/\lambda_L} +\frac{%
\lambda_L^2}{l_L^2}\rho^L_{20}e^{(x+d_1/2)/l_L},  \nonumber \\
M^L(x)&=&M^L_0(1-\frac{\lambda_L^2}{l_L^2})e^{(x+d_1/2)/l_L},\label{metal}
\end{eqnarray}
with similar solutions for the right hand side. In the
semiconductor, if $\lambda_N\ll x_0$,
\begin{eqnarray}
\rho^N(x)&=&\rho^{(1)}(x)+\rho^{(2)}(x),  \nonumber \\
\rho^{(1)}(x)&=&\frac{\lambda_N}{l_N}\rho^{(1)}_{10}e^{-(x-d_1/2)/\lambda_N} +\frac{%
\lambda_N^2}{l_N^2}\rho^{(1)}_{20}e^{-(x-d_1/2)/l_N},  \nonumber
\\
\rho^{(2)}(x)&=&\frac{\lambda_N}{l_N}\rho^{(2)}_{10}e^{(x-x_0+d_2/2)/\lambda_N}
\nonumber \\
&+&\frac{\lambda_N^2}{l_N^2}\rho^{(2)}_{20}e^{(x-x_0+d_2/2)/l_N}.\label{semi}
\end{eqnarray}
$M^{(1),(2)}$ can  be obtained similarly. All of coefficients in
(\ref{metal}) and (\ref{semi}) can be determined by using
Eqs.(\ref{1})-(\ref{4}) and the constraint (\ref{5}) and
(\ref{6}). The screening potential $W_0$ is determined by Gauss'
law. The total current is $j=\sum_s j_s$. Although $j_s$ are not a
constant, the total current $j$ is still a constant.

 \vspace{2mm}

\noindent{\it The spin-dependent current}: To demonstrate the
essential physics, we simplify matters by setting the parameters
of the metals and barrier widths on the left and right sides to be
the same: $\lambda_R=\lambda_L=\lambda$, $l_L=l_R=l$, $d_1=d_2=d$
and so on. The resistances of the barrier layers are taken as
$r^{(1)}=r^{(2)}=r$; $\gamma_1=\gamma_2=\gamma$ for the parallel
configuration and $\gamma_1=-\gamma_2=\gamma$ for the
anti-parallel configuration. To illustrate, we focus on the
calculation in the left barrier located at $x=0$. The tunneling
resistance is given by $ r_s^{(1)}=r(1-\gamma s)=
r_{0,s}\exp[d(\kappa_s(\mu)-\kappa_s(0))], $ where $
\kappa_s(\mu)\propto\int_0^ddx[2m(U-\Delta\mu_s(0)x/d)]^{1/2}, $
with $U$ the barrier height. That is $
\kappa_s(\mu)=\kappa_{0,s}\frac{2}{3\Delta\hat\mu_s(0)}[1-(1-
\Delta\hat\mu_s(0))^{3/2}]. $ Here $\Delta
\hat\mu_s=\Delta\mu_s/U$. The current  $j^L_s(x)$ at $x=0$ is
dependent on the bias voltage and the space charge distribution,
which is given by
\begin{eqnarray}
j_s^L(0)=A_sj_{0s},\label{current}
\end{eqnarray}
where
\begin{eqnarray}
A_s&=&1+\frac{4\pi\lambda^2}{l}\frac{\alpha(\beta-s)}{1-\alpha\beta}
\frac{\rho_{10}^L+M_0^L}{E},
\end{eqnarray}
and $j_{0s}=\sigma_s E$ is the current with no interaction;
$\beta$ ($\alpha$) measures the spin asymmetry of the
conductivities $\sigma_s$ (densities of states at the Fermi
surface $N_s$): $\sigma_s=\frac{\sigma}{2}(1+\beta s)$ and
$N_s=\frac 12N_F(1+s\alpha )$ where $N_F$ is related to the
screening length $\lambda$ by $\frac 1{N_F}=2\pi \lambda
^2\frac{1-\alpha ^2}{1-\alpha \beta }$. Noting that both
$\rho^L_{10}$ and  $M_0^L$  are proportional to the external
electric field $E$, $A_s$ is solely determined by the material
parameters.

Eq. (\ref{current}) implies that the spin-dependent current
$j_s(0)$ passing the interface differs a factor $A_s$ from the
non-interacting current $j_{0s}(0)$ which, for the parallel
configuration, is given by
\begin{eqnarray}
j^{s}_{0s}(0)\approx \frac{ V}{R_N+2r_{0,s}Y_s(0)},
\end{eqnarray}
where
$
Y_s(0)=e^{\kappa_{0s
}d[\frac{2}{3\Delta\hat\mu_s(0)}(1-(1-\Delta\hat\mu_s(0))^{3/2})-1]}.
$ For the anti-parallel configuration,
\begin{eqnarray}
&&j^a_{0s}(0)\approx\frac{V}{R_N+\sum_sr_{0,s}Y_s(0)},
\end{eqnarray}
which is almost spin-independent.

\vspace{2mm}

{\noindent \it The magnetoresistance:} Since the total current is
constant everywhere, we have, for the parallel configuration
\begin{eqnarray}
\frac{1}{R_P}=\sum_sj_s(0)/V
=\sum_s\frac{A_s^P}{R_N+2r_{0,s}Y^{P}_s(0)},
\end{eqnarray}
and for the anti-parallel configuration, we have
\begin{eqnarray}
\frac{1}{R_{AP}}=\sum_sj_s(0)/V
=\frac{2}{R_N+\sum_sr_{0,s}Y^{AP}_s(0)}.
\end{eqnarray}
From these, we obtain the magnetoresistance ratio
\begin{eqnarray}
&&\frac{\Delta
R}{R}\equiv\frac{R_{AP}-R_P}{(R_{AP}+R_P)/2}=\frac{2X}{2+X},\label{int}
\\
&&X=\sum_s\frac{A^P_s(R_N+\sum_{s'}r_{0,s'} Y^{AP}_{s'}(0))}
{2(R_N+2r_{0,s}Y^P_s(0))}-1.\nonumber
\end{eqnarray}
For non-interacting electrons, $A^P_+=A^P_-=1$ and one can see
that $\Delta R/R$ will not be beyond a maximal value (at $V\to 0$)
about 3.23\% for $r_{0+}:r_{0-}:R_N=1:2:1$ and decays as the bias
voltage increases. In Fig. 2, we plot the bias-dependence of the
magnetoresistance ratio for both the non-interacting and the
interacting case with $x_0\sim 1.25\mu$m. Here we take
$\lambda=0.1$nm, $l=20$nm, $\alpha=\beta=1/2$ and $l_N=3\mu$m
\cite{ab}. After the space charge effect is included, the ratio
(\ref{int}) grows as $\lambda_N$ becomes smaller. This screening
length can be controlled by doping the semiconductor (or by
applying an external gate voltage). The small $\lambda_N$ requires
heavy doping of the semiconductor. For a heavily doped
semiconductor, it may arrive at the giant magnetoresistance value,
say 56\% as shown in the figure, which raises orders of the
magnitude comparing to the magnetoresistance for the
non-interacting electrons.

There are some points to be emphasized in our calculations. i) In
the large $R_N$ limit, $\Delta R/R$ tends to zero no matter if
there is any interaction effect. ii) At zero bias, the
magnetoresistance is highest. In Fig. 2, we  show the dependence
of the magnetoresistance ratio on $\lambda_N/x_0$ for different
$x_0(\gg l)$ at zero bias \cite{note}. For a given $x_0$, we see
the magnetoresistance is also dependent on $l$. iii) The
magnetoresistance is reduced as the bias voltage increases. We
expect that the effective barrier height $U$, say several voltage,
for the FIS interface to be larger than that for the F-I-F
structure because the work function of the semiconductor is
smaller than that of a metal. So on an absolute scale of voltage,
this effect is not as big as in the conventional F-I-F structure.
Namely, the large magnetoresistance can be anticipated at an
experimental relevant bias voltage.

\vspace{2mm}

{\noindent \it The polarization:} We consider another experimental
measurable quantity \cite{STMST}, the polarization of current,
which is defined by $ P=\frac{j_+(0)-j_-(0)}{j_+(0)+j_-(0)}. $ We
first calculate $P$ by ignoring the space charge effect. In the
anti-parallel configuration $P$ is nearly zero. The polarization
$P$ for the parallel configuration is also small if $R_N$
dominates but it may arrive a considerable magnitude if the
$r_{0s}$ is of the same order of $R_N$, e.g., if the ratio
$r_{0+}:r_{0-}:R_N =1:2:1$, it may arrive at 40\% at the zero
bias. (In a wide range, our result is not very sensitive to this
ratio if these resistances have the same order of magnitude.) $P$
decays as the bias voltage increases, say for $V/U\sim 1$, $P$
drops to less than $10\%$.

To see the space charge effect, we look at $A_s$. We found that
$A_s$ are nearly independent of $l_N$ if $l_N$ is the longest
length scale. We fix $l_N=3\mu$m and $\kappa_{0s}d=300$. For the
anti-parallel configuration, the polarization is given by
$P^{ap}=\frac{A_+-A_-}{A_++A_-}$, which is bias-independent. In a
large screening length $\lambda_N$ and $x_0$, e.g, $100$nm and 1.5
$\mu$m, $P^{ap}$ is of the same order of the magnitude as that in
the non-interacting case. However,  for a small $\lambda_N$ and
$x_0$, e.g., 30 nm and $0.5\mu$m, $P^{ap}$ can be larger than one
because $j_+$, $j_-$ can go in opposite directions. $P^{ap}$
becomes significant when $\lambda_N$ and $x_0$ are getting
smaller. The physics behind this phenomenon can not be
well-understood yet at this stage. An intuitive consideration is
as follows: The effective driving potential $\Delta \mu_s$ is
different in sign between spin up and spin down electrons, because
the net charge density at the interface is less than the
magnetization density \cite{chui}. This inequality can be traced
to stem from the fact that the magnetic length is much larger than
the screening length. Because of this sign difference of the
driving potential, the currents for the two spin components may go
in opposite directions. However, this consideration has to be
examined thoroughly and we shall leave it to the further works.
The polarization of the parallel configuration is bias dependent.
For a large $\lambda_N$ and small $x_0$ (say 100 nm and
0.5$\mu$m), $P^p$ decays as the bias voltage increases. However,
for a small $\lambda_N$ (say 30 nm) and large $x_0$ (say 2$\mu$m),
$P^p$ increases as the bias voltage and finally saturates.
However, these bias dependent effects happen only in the high bias
regime, say $V/U>1$. Thus, for the practical usage, this bias
dependence may not be so important.

In conclusions, we have considered the space charge effect in a
multiple FIS junction. In the Hartree approximation, we found a
conspicuous difference of the spin-dependent current in the
parallel configuration from that in the non-interaction models
while the difference is minor for the anti-parallel configuration.
This leads to an order of magnitude growing in the
magnetoresistance ratio in such an approximation from the
non-interaction model if the parameters of the junction are
appropriately chosen. Although our theory is based on a specific
junction geometry and a simple approximation, it has indicated
that the electron interaction may play an important role in
nonequilibrium transport processes for some types of spintronic
devices on the spin injection. To quantitatively solve the
nonoequilibrium transport processes, a stricter treatment is
required.

This work was partially supported by the NSF of China. STC was
partially supported by the US NSF. He thanks the hospitality of
Institute of Theoretical Physics (Chinese Academy of Sciences,
Beijing) where this work was completed.

\vspace{2mm}

\noindent Fig. 1 The sketch of the geometry of the junction. The
quantities of the left (right) ferromagnet is indexed as 'L' ('R')
. \vspace{2mm}

\noindent Fig. 2 The magnetoresistances versus the bias voltage
for non-interacting junction and the junctions with the space
charge distribution. $\kappa_{0s} d=300$ has been taken.

\vspace{2mm}

\noindent Fig. 3 The magnetoresistance versus $\lambda_N/x_0$ in
$V=0$ for given semiconductor sizes $x_0$. The curves
corresponding to different $l$ are plotted in each diagram.


\begin{references}
\bibitem{ab} Throughout this paper, we take this set of the parameters
in all numerical results. For other values of $\alpha$ and
$\beta$, there will not be a qualitative change if they are not
very close to 0 or 1.
\bibitem{note} In the regime where $x_0 $ and $\lambda_N$ are of
the order of $l$, the situation is a bit subtle and requires  more
researches. Our investigation is limited in $\lambda_N\ll x_0$.
\end{references}
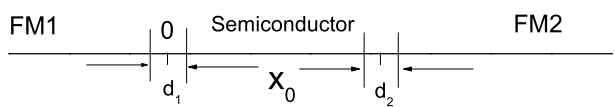
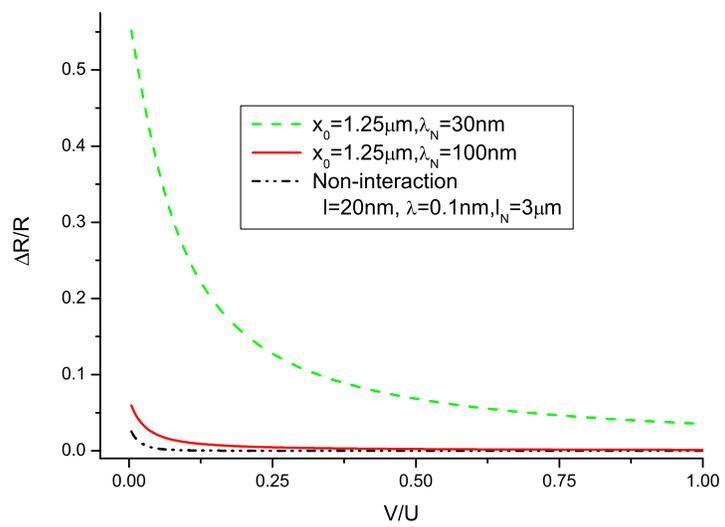
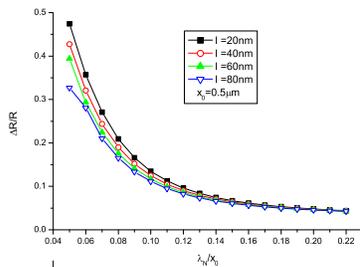
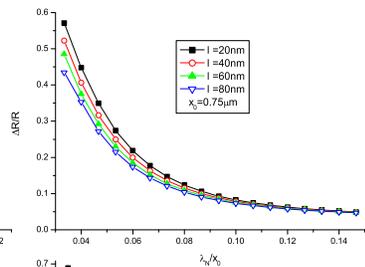
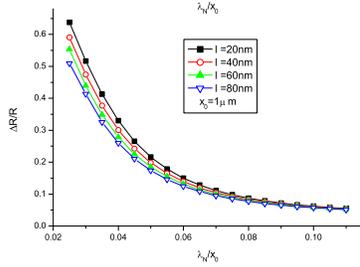
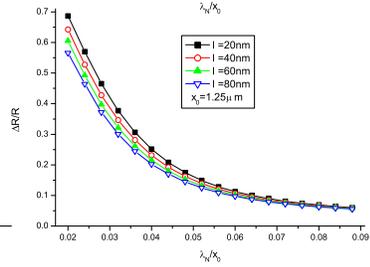
\end{document}